\documentclass[submission,copyright,creativecommons]{eptcs}
\providecommand{\event}{BEAT 2014} % Name of the event you are submitting to
\usepackage{breakurl}             % Not needed if you use pdflatex only.

\usepackage{amsmath}
\usepackage{amsfonts}
\usepackage{amssymb}
\usepackage{mathpartir}
\usepackage{enumerate}
\usepackage{color}
\usepackage{stmaryrd} % St Mary Road symbols for theoretical computer science. 
\usepackage{proof, prooftree}
\usepackage{tikz}
\usepackage{wrapfig}

\title{Dynamic Role Authorization in Multiparty Conversations}
\def\titlerunning{Dynamic Role Authorization in Multiparty Conversations} 
\author{
  Silvia Ghilezan %\inst{1}
    \institute{
  Univerzitet u Novom Sadu, Serbia}
  \and
  Svetlana Jak\v{s}i\'c %\inst{1}
  \institute{
  Univerzitet u Novom Sadu, Serbia}
    \and
  Jovanka Pantovi\'c %\inst{1}
  \institute{
  Univerzitet u Novom Sadu, Serbia}
  \and
  Jorge A. P\'erez %\inst{2}
  \institute{
    University of Groningen, The Netherlands}
  \and 
  Hugo Torres Vieira %\inst{3}
  \institute{
    LaSIGE, Faculdade de Ci\^encias, Universidade de Lisboa, Portugal}
}

\def\authorrunning{Ghilezan, Jak\v{s}i\'c, Pantovi\'c, P\'{e}rez, Vieira}

\newcommand{\til}[1]{\widetilde{#1}}

\newcommand{\NA}{a}
\newcommand{\NB}{b}
\newcommand{\NC}{c}

\newcommand{\RR}{r}
\newcommand{\RS}{s}

\newcommand{\RQ}{q}
\newcommand{\role}{\rho}
\newcommand{\roleS}{\sigma}

\newcommand{\Rauth}[1]{{\lceil #1 \rceil}}
\newcommand{\Runauth}[1]{{\lfloor #1 \rfloor}}
\newcommand{\Rnunauth}[1]{{\lfloor #1 \rceil}}

\newcommand{\PP}{P}
\newcommand{\PQ}{Q}
\newcommand{\PR}{R}
\newcommand{\inact}{0}
\newcommand{\parop}{\;|\;}
\newcommand{\rest}[1]{(\nu #1)}
\newcommand{\prefix}{\alpha}
\newcommand{\msg}{l}
\newcommand{\send}[4]{#1_#2!#3(#4)}
\newcommand{\receive}[4]{#1_#2?#3(#4)}
\newcommand{\sauth}[4]{#1_#2!#3(#4)}
\newcommand{\rauth}[4]{#1_#2?#3(#4)}
\newcommand{\sep}{\;\; \mid \;\;}

\newcommand{\red}{\rightarrow}
\newcommand{\subst}[2]{\{#1/#2\}}
\newcommand{\chsubst}[3]{\{#1:#2/#3\}}

\newcommand{\B}{B}
\newcommand{\T}{T}

\newcommand{\rname}[1]{{[#1]}}
\newcommand\roles[1]{{\tt pr}_2(#1)}
\newcommand\channels[1]{{\tt pr}_1(#1)}
\newcommand\funauth[2]{{\tt unauth}(#1,#2)}

\newcommand\bin[3]{?{#1}\,#2(#3)}
\newcommand\bout[3]{!{#1}\,#2(#3)}
\newcommand\bend{{\tt end}}
\newcommand\btau[2]{(#1\to #2):}
\newcommand\p{{\tt p}}

\newcommand{\bbtau}[4]{(#1\to #2): #3(#4)}

\newtheorem{example}{Example}

 \definecolor{darkgreen}{rgb}{0.0, 0.6, 0}
\newcommand\rr{{\color{darkgreen}{{\tt reviewer}}}}
\newcommand\rp{{\color{darkgreen}{{\tt professor}}}}
\newcommand\rs{{\color{darkgreen}{{\tt student}}}}
\newcommand\re{{\color{darkgreen}{{\tt editor}}}}
\newcommand\rpa{{\color{darkgreen}{\Rauth{\tt professor}}}}
\newcommand\rsa{{\color{darkgreen}{\Rauth{\tt student}}}}
\newcommand\rra{{\color{darkgreen}{\Rauth{\tt reviewer}}}}
\newcommand\rea{{\color{darkgreen}{\Rauth{\tt editor}}}}

\newcommand\rru{{\color{darkgreen}{\Runauth{\tt reviewer}}}}

\newcommand{\lp}{{paper}}

\newcommand{\lex}{{extend}}
\newcommand{\lf}{{final}}
\newcommand{\lread}{{read}}
\newcommand{\lreport}{{report}}
\newcommand{\lr}{{auth1}}
\newcommand{\lrr}{{auth2}}

\newcommand{\chr}{{\tt subm}}
\newcommand{\chj}{{\tt journal}}
\newcommand{\cha}{{\tt assist}}

\newtheorem{proposition}{Proposition}

\newtheorem{theorem}{Theorem}
\newtheorem{corollary}{Corollary}
\newtheorem{definition}{Definition}

\begin{document}

\maketitle

\begin{abstract}
Protocol specifications often identify the \emph{roles} involved in communications.
%typically for security purposes. 
In multiparty protocols that involve task delegation 
%Although such roles may be associated to actors or sites, 
it is often useful
to consider settings in which different sites may act on behalf of a single role. %; this is a recurring requirement 
%this is sensible in scenarios in which multiple parties collaborate on a task and when tasks may be delegated among participants. 
%In such settings,
It is then crucial to %devise mechanisms to 
control the roles that the different parties are authorized to represent, 
including the case in which role authorizations are determined only at runtime.
%allowing for authorizations to be dynamically passed along in messages. 
Building on previous work 
on conversation types 
with flexible role assignment, 
%that addresses the flexible assignment of roles to parties, 
here we report initial results on a typed framework for the analysis of multiparty communications with 
dynamic role authorization and delegation.
In the underlying 
process model, %is a simple $\pi$-calculus 
%in which 
communication prefixes are annotated with role authorizations 
and %in which 
authorizations can be passed around. 
%and by adding the possibility of 
%communicating authorizations in specially purposed primitives. We then 
We
extend the conversation type
system so as to statically distinguish processes that never incur in authorization errors.
The proposed static discipline guarantees that processes 
are always authorized to communicate on behalf of an intended role, also covering the case in which authorizations are dynamically
passed around in messages.

%We enhance a typed framework based on conversation types
%with the ability of exchanging 
%\emph{authorized roles}.
\end{abstract}

% !TEX root = main.tex
%========================
\section{Introduction}
%=========================
Different concepts of role-based performance 
%are widely accepted tools 
can be found in modern distributed information systems, ranging from access control to structured interactions in communication-centred systems. These concepts are typically grounded on the assumption that distinct participants (e.g., users %or terminals 
at different physical locations) may belong to the same role, and that a single participant may belong to (or implement) several different roles. Each role is associated with a set of permissions (e.g., privileges to access data or perform some action), thus enforcing an assignment of permissions to involved participants. In the case of multi-party interactions, a participant can use a role for communication only if the role is authorized for the particular action. 
As an example, consider the scenario of an 
electronic submission system, in which 
(confidential) paper submissions should be available only to 
only authorized participants. 
In this scenario,  editors typically rely on other participants who may act as reviewers. 
As such, any %potential reviewer 
participant
should be aware of the possibility of being appointed as reviewer. Also, 
a participant
 should be able to act as a reviewer only when she is officially authorized by the editor---this means, in particular, that unauthorized participants must not be able to read a submission. Furthermore, the exchanges determining an authorization should be part of the predefined protocols between the editor and the reviewer-to-be. 
%In that case, we have to take care about two important issues. First, both the reviewer and the subreviewer have to be aware of the possibility of engaging a subreviewer, and second, the subreviewer can do the subreviewing only if the reviewer authorizes him to do that.
%
%

In this paper, we consider the issue of enhancing multiparty communications with \emph{dynamic role authorizations}.
Our starting point is the typed framework given in~\cite{BaltazarCVV12}, based on \emph{conversation types}, 
%We consider this problem in the framework of dynamic roles in multiparty communication systems of . In \cite{BaltazarCVV12}, 
in which roles are flexibly assigned to participants that can act on their behalf.
While expressive, the model in \cite{BaltazarCVV12} does not check whether a given assigned role is indeed authorized to perform a particular action. The model that we propose here addresses this shortcoming, explicitly tracking the 
presence of (un)authorized  actions. Our typed model also enables the exchange of  authorizations along communication actions. Hence, participants may dynamically obtain authorizations to act on behalf of a role. We view our contribution as a first step in modeling and analyzing dynamic role-based communication and authorization, 
focusing on the identification of the basic ingredients that should be added on top of an existing framework in order to address the problem.

We present the main highlights of our model %through an example which formalizes 
by formalizing
the submission system.
Let us assume %that we are given 
the set of roles 
$\{\rp,$ $\rs,$ $\rr,$ $\re\}$ 
and the following 
global specification:
\[
\begin{array}[b]{l}
  \bbtau\re\rp \lr\rr.\\
  \quad\bbtau\rp\rs\lrr\rr.\\
  \qquad\bbtau\re\rr\lex{}.\\
  \quad\qquad\bbtau\rs\rp\lreport{}.\\
  \qquad\qquad\bbtau\rr\re\lf{}.\bend.
\end{array}
\]
%
%\begin{wrapfigure}[9]{r}{0.4 \linewidth}
%%\vskip-1ex
%\begin{tikzpicture}[semithick]
%  \draw [thick] (0,0)--(0,4);
%  \draw [thick] (3,0)--(3,4);
%  \draw [thick] (6,0)--(6,4);
%  \draw [dotted,->,blue] (0,3.5)--(3,3.5);
%  \node [blue] at (1.3,3.7) {\em paper};
%  \draw [->] (0,3)--(3,3);
%  \node at (1.4,3.2) {\em auth1({\tt reviewer})};
%%  \draw [->] (0,3)--(3,43;
%%  \node at (1.3,3.2) {\em instruction};
%  \draw [dotted,->,blue] (3,2.5)--(6,2.5);
%  \node [blue] at (4.5,2.7) {\em read};
%  \draw [->] (3,2)--(6,2);
%  \node at (4.5,2.2) {\em auth2({\tt reviewer})};
%  \draw [<-] (6,1.5)--(0,1.5);
%  \node at (1.3,1.7) {\em extension};
%  \draw[->] (6,1)--(3,1);
%  \node at (4.3,1.2) {\em report};
%  \draw[->] (3,0.5)--(0,0.5);
%  \node at (1.3,0.7) {\em final};
%\end{tikzpicture}
%\end{wrapfigure}
%=======old text
%\noindent In this example, the $\re$ is allowed to sending authorizations for the role $\rr$.
%The global specification says that the $\re$ authorizes the $\rp$ to act as the $\rr.$
%The $\rp$ asks a PhD $\rs$ to read the paper and passes him the $\rr$ role for that purpose.
%%Afterwards, $\re$ approves request for extension of the deadline, awaits to receive the final decision from the $\rr$ and finally take back the authorization from the $\rp$ to act as the $\rr.$  
%The PhD student, now acting as the $\rr,$ gets the deadline extension from the $\re.$ As $\rs,$ he sends the report.  After the $\rp$ receives the report, he sends his final decision to the $\re$. % and gives back the $\rr$ role.
%====================
\noindent %In this example, t
Above, role $\re$ is allowed to send authorizations for the role $\rr$.
The global specification says that the $\re$ authorizes the $\rp$ to act as  $\rr,$ which is  followed by passing the authorization for the role $\rr$ from the $\rp$ to the $\rs.$
%Afterwards, $\re$ approves request for extension of the deadline, awaits to receive the final decision from the $\rr$ and finally take back the authorization from the $\rp$ to act as the $\rr.$  
%The PhD student, now acting as 
The $\rr$ gets a deadline extension from the $\re,$ then $\rs$ sends the report to the $\rp$. Finally, the $\rr$ sends a final decision to the $\re$. % and gives back the $\rr$ role.
%One possible implementation of this protocol in our process model is the following 
We may implement this specification as the process 
  $$Sys = \rest{\chr}\send{\chj}{\rea}{\lp}{\chr}.\PP'\;\mid\;\receive\chj\rpa\lp\NA.\PQ'\; \mid \; \PR$$
where
$\chr$ denotes a channel and processes $\PP'$, $\PQ'$, and $\PR$ are defined as:\\
  $\PP' = \sauth\chr\rea\lr\rra.\PP''\\
  \PP''=
  \send\chr\rea\lex{}. 
  \receive\chr\rea\lf{}.\inact \\
  \PQ' = \rauth\NA\rpa\lr\rr.\PQ''\\
  \PQ''=
  \send\cha\rpa\lread\NA.
  \sauth\NA\rpa\lrr\rru.\\
  \hspace*{8.5mm} \receive\NA\rpa\lreport{}.
  \send\NA\rru\lf{}.\inact \\
  \PR = \receive\cha\rsa\lread\NB.\rauth\NB\rsa\lrr\rr.
  \receive\NB\rru\lex{}.
  \send\NB\rsa\lreport{}.\inact$\\\\
In our process model, each communication prefix $\prefix$ is decorated with either $\Rauth{r}$ 
(i.e., role $r$ is authorized to perform $\prefix$) 
or $\Runauth{r}$ (i.e., role $r$ is \emph{not} authorized to perform $\prefix$).
%Given a role $r$, we write $\Rauth{r}$ and $\Runauth{r}$ to denote the fact that $r$ is authorized and unauthorized for a given process prefix, respectively. 
These decorations define fine-grained specifications of (un)authorized communication actions. 
The  three subprocesses of $Sys$ formalize the behavior of the editor, the professor, and the student, respectively.
%There are three key processes, one initially authorized as $\re$ and $\rr,$ one as $\rp,$ and the last one as $\rs.$
%These auhtorized roles are specified by $\rra,$ $\rea,$ $\rpa$ and $\rsa.$
The first subprocess creates a fresh channel $\chr$ which is passed (on behalf of $\re$) to the second subprocess (that receives it as $\rp$) over the message $\lp$ on  channel $\chj.$ 
%By this interaction, the given process reduces to the process
%We thus have the following process reduction:
Process $Sys$ then reduces to 
%\centerline{
$$\rest{\chr}(\sauth\chr\rea\lr\rra.\PP''\;\mid\;\rauth\chr\rpa\lr\rr.\PQ''\subst{\chr}{\NA}\;\mid\;\PR)$$
%}
Here the process authorized as $\re$ sends authorization for the role $\rr$  to the process acting on behalf of $\rp.$
After interaction, the role $\rr$ will become authorized in $\PQ''\subst{\chr}{\NA},$ so the second subprocess will reduce to 
\begin{center}
$
  \send\cha\rpa\lread\chr.
  \sauth\chr\rpa\lrr\rra.$\\
  $\hspace*{8.5mm}\receive\chr\rpa\lreport{}.
  \send\chr\rra\lf{}.\inact
$
\end{center}
Continuing along these lines, process $\PR$ joins the conversation on channel $\chr,$ gets authorization for $\rr,$ receives $\lex$ as $\rr,$ sends $\lreport$ as $\rs,$ and finally the second subprocess sends $\lf$ decision, as $\rr,$ to the process acting as $\re.$
It is worth observing that the initial specification for the student (cf. process $\PR$) is authorized to act as $\rs$ but it is not authorized to act as $\rr$. The required authorization to access and review the submission should result as a consequence of an interaction with the process realizing the behavior of the professor. That is, previous communication actions directly determine current authorization privileges for interacting partners. As such, the issue of ensuring consistent conversations is tightly related to issues of role authorization and deauthorization. To address this combination of issues, the type discipline that we present here ensures that structured multiparty conversations are consistent with respect to both global protocols and requirements of dynamic role authorization (cf. Corollary~\ref{c:safety}).

This  paper is organized as follows. In \S\,\ref{sec:language}, we define our process language and illustrate further the intended model of dynamic role authorization. \S\,\ref{sec:types} presents the type system and the properties for well-typed processes. Finally, in \S\,\ref{sec:related}, we comment on related works and discuss open problems.
%\emph{For reviewer's convenience, the appendix collects auxiliary definitions and proofs.}
%!TEX root =  main.tex
\section{Process Language}
\label{sec:language}
\paragraph{Syntax.}

We consider a %variant the 
synchronous $\pi$-calculus~\cite{SangiorgiW01} extended with labelled communications and prefixes for authorization 
sending and receiving. 
Let $\mathcal{L},\mathcal{R}$, and $\mathcal{N}$ be infinite base sets of \emph{labels}, \emph{roles}, and \emph{channels}, respectively.
We use $\msg, \ldots$ to range over $\mathcal{L}$; $\RR, \RS, \ldots$ to range over $\mathcal{R}$; and $\NA, \NB, \NC, \ldots$ to range over $\mathcal{N}.$ 
%In our model, a 
A role $\RR$ can be qualified as \emph{authorized} (denoted $\Rauth\RR$) or as \emph{unauthorized} (denoted $\Runauth\RR$). 
We write  $\Rnunauth\RR$ to denote a role $r$ with some qualification, 
and use 
$\role, \roleS$ to range over $\Rnunauth\RR$ for some unspecified $r$.

As motivated above, each communication prefix in our calculus is decorated by a qualified role for performing the associated action.
Intuitively, prior to execution not all roles have to be  authorized; 
we expect unauthorized roles may have the potential of becoming authorized as the structured interactions take place.
In fact, we expect all actions of the system to be associated to authorized roles; top-level prefixes on unauthorized roles are regarded as \emph{errors}.
To enable dynamic role authorization, our model allows for the exchange of the authorization on a role. 
%More formally, t
Formally, the syntax of processes is given by: 
%
%%\begin{table}
%\begin{math}
%\displaystyle
%\begin{array}[t]{@{}rcl@{\quad}l@{}}
%  \PP,\PQ & ::= & \inact & \text{(Inaction)} \\
%          & | & \PP\parop\PQ & \text{(Parallel)}\\
%          & | & \rest\NA P & \text{(Restriction)}\\
%          & | & \prefix.\PP & \text{(Prefix)}
%\end{array}
%\qquad
%\begin{array}[t]{@{}rcl@{\quad}l@{}}
%  \prefix   & ::= & \send\NA\role\msg\NB & \text{(Output)} \\
%            &  |  & \receive\NA\role\msg\NB & \text{(Input)}\\
%            &  |  & \sauth\NA\role\msg\roleS & \text{(Send authorization)}\\
%            &  |  & \rauth\NA\role\msg\RR & \text{(Receive authorization)}\\\\
%
%  \role   & ::= & \Rauth\RR & \text{(Authorized role)} \\
%          &  |  & \Runauth\RR & \text{(Unauthorized role)} \\\\
%\end{array}
%\end{math}
%%\caption{\label{tab:syntax}Syntax of processes}
%%\end{table}
%
\begin{eqnarray*}
  \PP,\PQ   & ::= & \inact \quad | \quad \PP\parop\PQ  \quad | \quad \rest\NA P \quad | \quad \prefix.\PP \\
  \prefix   & ::= & \send\NA\role\msg\NB  \quad | \quad \receive\NA\role\msg\NB \quad | \quad \sauth\NA\role\msg\roleS \quad | \quad \rauth\NA\role\msg\RR
  \qquad \qquad
  \role      ::=  \Rauth\RR \; | \; \Runauth\RR 
\end{eqnarray*}
Constructs for 
inaction ($\inact$), parallel composition ($\PP\parop\PQ$), and restriction ($\rest\NA P$) are standard. %usual $\pi$-calculus operators.
We write $(\nu\NA_1, \ldots, \NA_n)$ as a shorthand for $\rest{\NA_1}\cdots \rest{\NA_n}.$
Also, we write $\til{a}$ to stand for the sequence of names $\NA_1, \ldots, \NA_n$.
To define communication of channels and authorizations, our language has four kinds of prefixes, denoted $\prefix$. Each prefix is associated to a $\role$. Intuitively, a prefix associated to $\Rauth\RR,$ is said to be \emph{authorized} to perform the associated action  under role $r$. 
A prefix associated to $\Runauth\RR$ is not authorized to perform the corresponding action as $r$; but it may be the case that such prefixes are
dynamically authorized via communication. 
%The first pair of input/output processes consists of:
The intuitive semantics for prefixes follows:
\begin{enumerate}[-]
\item $\send\NA\role\msg\NB$ expresses sending of name $\NB,$ in labelled message $\msg,$ along channel $\NA,$ under qualified role $\role$;  
\item $\receive\NA\role\msg\NB$ expresses receiving of name $\NB,$ in labelled message $\msg,$ along channel $\NA,$ under qualified role $\role.$
\end{enumerate}
These two prefixes are taken from~\cite{BaltazarCVV12}, here extended in with authorization control via 
role qualification.
% We aim to allow such actions only to the roles that are given such an authorization.
The second pair of prefixes is new to our calculus:
\begin{enumerate}[-]
\item $\sauth\NA\role\msg\roleS$ expresses sending of the \emph{qualified role} $\roleS,$ in labelled message $\msg,$ along channel $\NA,$ under  qualified role $\role$;
\item $\rauth\NA\role\msg\RR$ expresses receiving authorization for role $\RR,$ in labelled message $\msg,$ along channel $\NA,$ under  qualified role $\role.$
\end{enumerate}
%
%We will say that a prefixed process is an $(\NA,\msg)$-process if the prefix has the subject $\NA$ and the label $\msg.$ For all $\NA\in\mathcal{N}$ and $\msg\in\mathcal{L},$ an $(\NA,\msg)$-redex is one of the following pairs of $(\NA,\msg)$-processes:
%\begin{itemize}
%  \item[(i)]  $\send\NA\role\msg\NB$ and $\receive\NA\roleS\msg\NC,$ for any $\NB,\NC\in\mathcal{N}$ and any qualified roles $\role$ and $\roleS,$ and 
%  \item[(ii)] $\sauth\NA\roleS\msg{\Rnunauth\RR}$ and $\rauth\NA\roleZ\msg\RR$ for any $\RR \in \mathcal{R},$ and any qualified roles $\roleS$ and $\roleZ.$
%\end{itemize}
%
%If we need to take into account the four arguments of the prefix $\prefix,$ we will write $\alpha(\NA,\msg,\role,\varv),$ for $\varv\in\{\NB,\role,\RR\}.$

\paragraph{Operational Semantics.}

%\begin{table}
%\hline
%\[
%\begin{array}{@{}c@{}} 
%  \inferrule[]{}
%  {\PP\parop\inact\equiv\PP}\qquad
%  \inferrule[]{}
%  {\PP\parop\PQ\equiv\PQ\parop\PP}\qquad
%  \inferrule[]{}
%  {(\PP\parop\PQ)\parop\PR\equiv\PP\parop(\PQ\parop\PR)}\qquad
%  \inferrule[]{}
%  {\rest\NA\inact\equiv\inact}\\
%  \inferrule[]{}
%  {\rest\NA\rest\NB\PP\equiv\rest\NB\rest\NA\PP}\qquad
%  \inferrule[]{}
%  {\PP\parop\rest\NA\PQ\equiv\rest\NA(\PP\parop\PQ)\quad\text{if $\NA\notin\fn\PP$}}
%\end{array}
%\]
%\caption{\label{tab:structural}Structural congruence}
%\end{table}
%Given properties are the same as in the standard $\pi$-calculus.
\begin{figure}[t!]
\[
\begin{array}{@{}c@{}} %\hline \\
  \inferrule[\rname{R-Comm}]
  {}
  {\send\NA{\Rauth\RS}\msg\NB.\PP\parop\receive\NA{\Rauth\RR}\msg\NC.\PQ
  \red
  \PP\parop\PQ\subst\NB\NC} 
  \qquad
%  \inferrule[R-Authorization]
%  {{\color{red}\PP'=\deauthorize\PP\NA\RQ}\quad\PQ'=\authorize\PQ\NA\RQ}
%  {\sauth\NA{\Rauth\RS}\msg{\Rauth\RQ}.\PP\parop\rauth\NA{\Rauth\RR}\msg\RQ.\PQ
%  \red
%  \PP'\parop\PQ'}
%  \\\\
  \inferrule[\rname{R-Auth}]
  {}
  {\sauth\NA{\Rauth\RS}\msg{\Rauth\RQ}.\PP\parop\rauth\NA{\Rauth\RR}\msg\RQ.\PQ
  \red
  \PP \parop\PQ\chsubst{\NA}{\Rauth\RQ}{\Runauth\RQ}} 
  \vspace{2mm}\\
  \inferrule[\rname{R-Par}]
  {}{\PP\red\PP' ~\Rightarrow~ \PP\parop\PQ\red\PP'\parop\PQ}
  %{\PP\red\PP'}
  %{\PP\parop\PQ\red\PP'\parop\PQ}
  \qquad
  \inferrule[\rname{R-Restriction}]
  {}{\PP\red\PQ ~\Rightarrow~ \rest\NA\PP\red\rest\NA\PQ}
  %{\PP\red\PQ}
  %{\rest\NA\PP\red\rest\NA\PQ}
  \qquad
  \inferrule[\rname{R-Struct}]
  {}{\PP\equiv\PP', \PP'\red\PQ', \PQ'\equiv\PQ ~\Rightarrow~ \PP\red\PQ}
  %{\PP\equiv\PP'\quad\PP'\red\PQ'\quad\PQ'\equiv\PQ}
  %{\PP\red\PQ}\\ % \\ \hline
\end{array}
\]
\caption{\label{tab:reduction}\strut Reduction relation}
\end{figure}

%\vspace{1cm}
%\begin{math}
%\begin{array}{lcl}
%\authorize\inact\NA\RR         & = & \inact\\
%\authorize{\PP\parop\PQ}\NA\RR & = & \authorize\PP\NA\RR\parop\authorize\PQ\NA\RR\\
%\authorize{\rest\NA\PP}\NB\RR  & = & \rest\NA\authorize\PP\NB\RR \\
%\authorize{\prefix.\PP}\NA\RR  & = & \left\lbrace 
%                                     \begin{array}{ll}
%                                     \send\NA{\Rauth\RR}\msg\NB.\authorize\PP\NA\RR, & \text{if } \prefix=\send\NA{\Runauth\RR}\msg\NB\\
%                                     \receive\NA{\Rauth\RR}\msg\NB.\authorize\PP\NA\RR, & \text{if } \prefix=\receive\NA{\Runauth\RR}\msg\NB\\
%                                       \sauth\NA{\Rauth\RR}\msg{\Rauth\RS}.\authorize\PP\NA\RR, & \text{if } \prefix=\sauth\NA{\Runauth\RR}\msg{\Rauth\RS}\\
%                                     \sauth\NA{\Rauth\RS}\msg{\Rauth\RR}.\authorize\PP\NA\RR, & \text{if } \prefix=\sauth\NA{\Rauth\RS}\msg{\Runauth\RR}\\
%                                     {\color{green}{\sauth\NA{\Runauth\RS}\msg{\Rauth\RR}.\authorize\PP\NA\RR,}} & {\color{green}{\text{if } \prefix=\sauth\NA{\Runauth\RS}\msg{\Runauth\RR}}}\\
%                                     \prefix.\authorize\PP\NA\RR, & \text{otherwise}
%                                     \end{array}                                                                          
%                                     \right. 
%\end{array}
%\end{math}

%$\NA:\subst{\sigma}{\overline{\sigma}} :=\subst{\NA_{\sigma}}{\NA_{\overline{\sigma}}}\subst{\sauth\NA{\rho}\msg{\sigma}}{\sauth\NA{\rho}\msg{\overline{\sigma}}}.$ 
The process semantics %of our process model 
is defined via 
a reduction relation, which is
defined in Figure~\ref{tab:reduction} and 
denoted $\red$. 
Reduction is closed under static contexts and 
%structural congruence and reduction relations between processes.
\emph{structural congruence}, denoted $\equiv$ and defined in standard lines (cf.~\cite{SangiorgiW01}).
%the smallest congruence that is closed under alpha-conversion and satisfies the axioms 
%in Figure~\ref{FigStruct} (given in the Appendix---essentially the same as in the $\pi$-calculus).
To support communication of qualified roles, we use a form of substitution denoted by $\NA:\subst{\sigma}{\overline{\sigma}}$, representing the substitution $\subst{\sigma}{\overline{\sigma}}$ applied only on channel $\NA.$ In turn, this includes two substitutions, for prefix subjects and qualified roles occurring as communication objects, respectively: $\subst{\NA_{\sigma}}{\NA_{\overline{\sigma}}}$ and $\subst{\sauth\NA{\rho}\msg{\sigma}}{\sauth\NA{\rho}\msg{\overline{\sigma}}}.$ In Figure~\ref{tab:reduction} rules \rname{R-Comm} and \rname{R-Auth} specify 
synchronizations: %possible interactions. By both rules, 
%In both rules, 
two processes can exchange a message 
($\msg$) on a channel ($\NA$) only under authorized roles (denoted $\Rauth\RS, \Rauth\RR$ in the rules). Using rule \rname{R-Auth}, a process can authorize another process to act under a role ($\RQ$) only if the first process has permission to do such an authorization ($\Rauth\RQ$) and the second process is asking for authorization of the same role ($\RQ$). 
%As the considered role will be unauthorised in the continuation of the first process, we consider this interaction as a kind of  authorization delegation. 
We use $\red^*$ to denote the reflexive and transitive closure of $\red.$
The following definitions are key to syntactically distinguish authorization errors:

%Jorge: A couple of new definitions. 

\begin{definition}[Unauthorized Prefix / Errors] Let  $\alpha$ and $P$ be a prefix and a process as defined above.
\begin{enumerate}[-]
\item We say $\alpha$ is   \emph{unauthorized} if its subject is associated to an unauthorized role or 
its output object is an unauthorized role, i.e.,  if $\alpha = \receive{\NA}{\Runauth\RR}\msg{\NB}{}$, 
$\alpha = \send{\NA}{\Runauth\RR}\msg{\NB}{}$, $\alpha=\receive{\NA}{\Runauth\RR}\msg{\RS}{},$ $\alpha=\send{\NA}{\Runauth\RR}{\msg}{\Rauth{\RS}},$ or
$\send{\NA}{\sigma}{\msg}{\Runauth{\RS}}$, for some $\NA, \RR, \msg, \NB, \RS, \sigma$.
We write $\Runauth\alpha.P$ 
instead of $\alpha.P$
whenever $\alpha$ is unauthorized.
%\end{definition}

%We may then define authorization errors simply as the occurrence of unauthorized prefixes:
%The following notion defines errors with respect to unauthorized roles:

%\begin{definition}[Authorization Error]
\item We say that $\PP$ is an \emph{authorization error}
if $\PP\equiv (\nu \widetilde{\NA}) (\Runauth\alpha.\PQ\parop \PR)$, for some $\widetilde{\NA}, \alpha, \PQ, \PR$.
\end{enumerate}
\end{definition}

Notice that an authorization error is a ``stuck process'' according to our semantics, i.e., a process which cannot synchronize since it does not have the required authorization to do so.

\begin{example}
To illustrate reduction and authorization errors, consider processes $P$ and $Q$ below:
\begin{eqnarray*}
    P & = & \receive{\NA}{\Rauth\RR}{\msg_2}\NC.\receive\NC{\Rauth\RR}{\msg_1}\RS.\receive{\NC}{\Runauth\RS}{\msg_3}{ }.0\parop
    \rest\NB\send{\NA}{\Rauth\RQ}{\msg_2}{\NB}.\send\NB{\Rauth\RQ}{\msg_1}{\Rauth{\RS}}.\send\NB{\Rauth\RQ}{\msg_3}{}.0 \\
Q & = &        \rest\NB (\send\NB{\Rauth\RQ}{\msg_1}{\Rauth{\RS}}.\send{\NA}{\Rauth\RQ}{\msg_2}{\NB}.\send\NB{\Rauth\RQ}{\msg_3}{}.0 \parop
    \receive\NB{\Rauth\RR}{\msg_1}\RS.0) \parop
    \receive{\NA}{\Rauth\RS}{\msg_2}\NC.\receive{\NC}{\Runauth\RS}{\msg_3}{ }.0
\end{eqnarray*}
In both $P$ and $Q$, channel $b$ is used according to the specification 
  $\bbtau\RQ\RR{\msg_1}\RS.\bbtau\RQ\RS{\msg_3}{}$
  which informally says first there is an interaction from role $\RQ$ to role $\RR$ on message $\msg_1$, exchanging authorization on role $\RS$, followed
  by an interaction between $\RQ$ and $\RS$ on $\msg_3$ (where, for the sake of simplicity, we omit contents of the message).
  We may infer the following reductions for $P$:
 \begin{eqnarray*}
  P  & \red & 
    \rest\NB
    (\receive\NB{\Rauth\RR}{\msg_1}\RS.\receive{\NB}{\Runauth\RS}{\msg_3}{ }.0\parop
    \send\NB{\Rauth\RQ}{\msg_1}{\Rauth{\RS}}.\send\NB{\Rauth\RQ}{\msg_3}{}.0) %\\
     ~~\red~~  \rest\NB
    (\receive{\NB}{\Rauth\RS}{\msg_3}{ }.0\parop
    \send\NB{\Rauth\RQ}{\msg_3}{}.0)
  \end{eqnarray*}
and so % we can observe that 
all actions are carried out on behalf of authorized roles.
%, since the authorization passed in the synchronization allows the process on the lhs to dynamically obtain the authorization.
  In contrast, we have that 
    \[ Q \red^* \rest\NB(\send\NB{\Rauth\RQ}{\msg_3}{}.0 \parop \receive{\NB}{\Runauth\RS}{\msg_3}{ }.0).\]
    and so we infer that $Q$ is ill-behaved since it reduces to an authorization error on role $s$.
\end{example}

As the previous example illustrates, there are processes  which respect communication specifications but lead to authorization errors.
The type system described in the following section addresses this issue.

%\input{example3}
%!TEX root =  main.tex
%=========================
\section{Type System} \label{sec:types}
%=========================

We consider the conversation types language as presented in~\cite{BaltazarCVV12}, extending message type $M$ with the role $\RR$, 
so that we may capture role authorization passing. This is a rather natural extension, formally given by the syntax in Figure~\ref{tab:conversation_types}. Behavioral types $B$ 
include: % inaction 
$\bend$, which describes inaction; $B \parop B$, which allows to describe concurrent independent behavior; the sometime type $\Diamond B$, 
which says that behavior $B$ may take place immediately or later on. Finally, a behavioral type $\p \msg(M).B$ describes a communication action 
identifying the role or roles involved, and whether the action is an input $?{\RR}$ or an output $!{\RR}$ or a message exchange $\btau{\RR}{\RR}$, 
a carried message type $M$ and the behavior that is prescribed to take place after the communication action $B$. 

We use behavioral types to specify the interactions in \emph{linear} channels, where no communication races are allowed
(which is to say that at any given moment, there can only be one matching pair of input/output actions). 
In our setting, where several parties may simultaneously use a channel, this \emph{linear} communication pairing is ensured via message labels:
at a given moment, there can be only one pair of processes able to exchange a labelled message. 
For shared channels, where communication races are allowed, we ensure consistent usage (but no structured protocol of interaction) via
shared channel types $T,$ which carry a (linear) behavioral type describing the usage delegated in the communication.

Message type $M$ also captures the usages delegated in communications: in case $M$ is a behavioral type $B$ or a shared channel type $T$
it describes how the receiving process uses the received channel; in case $M$ identifies a  role $\RR$ then the message type captures an
authorization delegation in the specified role.

%\begin{figure}[t!]
%%\hline
%\[
%  \begin{array}{lcll} \hline \\
%        B   & ::=  & \; \bend \sep B \parop B  \sep \Diamond B  \sep \p \msg(M).B  & \mbox{ Behavioral types } \\
%        T   & ::=  & \; \msg(B)   & \mbox{ Shared channel types }\\
%        M   & ::=  & \; B \sep T \sep \RR & \mbox{ Message types }\\
%       \p   & ::=  & \; !{\RR} \; \sep ?{\RR}  \sep \btau{\RR}{\RR} & \mbox{ Communication prefix }\\\\ \hline
%\end{array}
%%\hline\hline
%\]
%\caption{Conversation types}
%\label{tab:conversation_types}
%\end{figure}
%
\begin{figure}[t!]
%\hline
\[
  \begin{array}{lcllcl} %\hline \\
        B   & ::=  & \; \bend \sep B \parop B  \sep \Diamond B  \sep \p \msg(M).B & \qquad   T   & ::=  & \; \msg(B)\\
        M   & ::=  & \; B \sep T \sep \RR                                         & \qquad \p   & ::=  & \; !{\RR} \; \sep ?{\RR}  \sep \btau{\RR}{\RR} \\ %\\ \hline
\end{array}
%\hline\hline
\]
\caption{Conversation types}
\label{tab:conversation_types}
\end{figure}

Type environments %$\Delta$ and $\Gamma$ 
separate linear and shared channel usages:  $\Delta$ associates channels with (linear) behavioral types, 
given by $ \Delta \;::=\; \emptyset \sep \Delta, \NA:\B$,
whereas $\Gamma$ associates channels with shared channel types, given by $\Gamma \; ::= \; \emptyset \sep \Gamma, \NA:\T$.
%\begin{eqnarray*}
%  \Delta & ::= & \emptyset \sep \Delta, \NA:\B \qquad \qquad \Gamma \; ::= \; \emptyset \sep \Gamma, \NA:\T
%\end{eqnarray*}
%Environments $\Delta$ and $\Gamma$ serve to describe interactions on linear and shared channels, respectively.
Typing rules rely on subtyping as well as on operators for apartness, well-formedness,  and splitting of types.
We refer to~\cite{BaltazarCVV12} for a details on these operations. %; Appendix~\ref{sec:typeops} collects supporting definitions.
While \emph{type apartness} ($\#$) refers to independent behaviors ensured via disjoint (message) label sets, \emph{well-formedness} %(cf. Figure~\ref{fig:WTypes}) 
ensures that parallel behaviors are apart and that the sometime $\Diamond$ is not associated to message
synchronizations --- synchronizations are not allowed to take place sometime later, they are always specified to take place at a given stage in 
the protocol. The \emph{subtyping relation} $<:$ %(cf. Figure~\ref{fig:subtype}), 
allows for (some) behaviors which are prescribed to take place 
immediately to be used in contexts that expect such behaviors to take place sometime ($\Diamond$) further along. 
Type \emph{splitting} %(given in Figure~\ref{fig:Split}) 
supports the distribution of protocol ``slices'' among the participants in a conversation: we write $B = B_1 \circ B_2$ to say 
that behavior $B$ may be split in behaviors $B_1$ and $B_2$ so that an overall behavior $B$ may be distributed (e.g., in the two branches of 
a parallel composition). We remark that $B_1$ and $B_2$ may be further  split so as to single out the individual contributions of each participant in a conversation, where decomposition is driven by the structure of the process in the typing rules. 

We lift the split relation to 
$\Delta$ type environments in unsurprising lines: $\Delta, \NA:\B = \Delta_1,\NA:\B_1 \circ \Delta_2,\NA: \B_2$ if
$\B = \B_1 \circ \B_2$ and $\Delta = \Delta_1 \circ \Delta_2$, and also $\Delta, \NA:\B = \Delta_1,\NA:\B \circ \Delta_2$ (and symmetrically) 
if $\Delta = \Delta_1 \circ \Delta_2$. In typing rules we write $\Delta_1 \circ \Delta_2$ to represent $\Delta$ (if there is such $\Delta$) such that
$\Delta = \Delta_1 \circ \Delta_2$.
%The definition of apartness is as in~\cite{BaltazarCVV12}. 

A typing judgment is of the form $\Delta; \Gamma \vdash_{\Sigma} \PP.$  The \emph{authorization set} $\Sigma$ is a subset of the direct product of 
the set of channel names and the set of roles, i.e.,
$\Sigma \subseteq \mathcal{N}\times \mathcal{R}$.
The typing judgment states that the process $\PP$ is well typed under $\Delta$ and $\Gamma$ with 
roles from $\roles\Sigma$ (the projection on the second element of the pairs in $\Sigma$) appearing in $\PP$ unauthorized on corresponding channels 
from $\channels\Sigma.$

\begin{figure}[t!]%[p]
\footnotesize{ 
%\hline
%\[
%%\rho\in\{\Rauth\RR,\Runauth\RR\}, \sigma\in\{\Rauth\RS,\Runauth\RS\}, 
%\text{Define: }
%\funauth\NA{\Runauth\RQ}=\{(\NA,\RQ)\},\funauth\NA{\Rauth\RQ}=\emptyset
%\]
\[	
\begin{array}{lcl}
%\hline\\
  \text{[T-end]}  && 
     \prooftree
     \justifies
	 \Delta_{\tt end}; \Gamma \vdash_{\emptyset}  0
	 \endprooftree
%  \\[8mm]
    \qquad \qquad
    \text{[T-snew]}  \qquad
     \prooftree
     \Delta; \Gamma, \NA:l(\B) \vdash_\Sigma \PP 
     \justifies
	 \Delta; \Gamma \vdash_{\Sigma} (\nu \NA) \PP
	 \endprooftree
  \\[8mm]
  \text{[T-new]}  && 
     \prooftree
     \Delta, \NA:B;\Gamma \vdash_{\Sigma} \PP \qquad {\tt matched}(B) \qquad \NA\not\in \channels{\Sigma}
     \justifies
	 \Delta; \Gamma \vdash_{\Sigma} (\nu \NA) \PP
	 \endprooftree
	 \qquad 
	   \text{[TProc-par]} \quad
	   \prooftree
	 \Delta_1; \Gamma \vdash_{\Sigma} \PP \qquad \Delta_2; \Gamma \vdash_{\Xi} \PQ
	\justifies
	\Delta_1 \circ \Delta_2; \Gamma \vdash_{\Sigma\cup\Xi} \PP \parop \PQ
		 \endprooftree 
  \\[8mm]
  \mbox{[Trole-in]}  && 
  \prooftree
	\Delta\circ \NA:B; \Gamma \vdash_{\Sigma \cup \{(\NA, \RS)\}} \PP \quad \bin\RR\msg\RS.B <: B' \quad  \Xi= \Sigma \cup \funauth{\NA}{\Rnunauth\RR}
	\justifies
	\Delta \circ \NA:B' ; \Gamma \vdash_{\Xi} \rauth \NA{\Rnunauth\RR}\msg\RS.\PP
	 \endprooftree
  \\[8mm]
  \mbox{[Trole-out]}  && 
  \prooftree
  {\begin{array}{l}
	\!\!\!\Delta \circ \NA:B; \Gamma \vdash_{\Sigma} \PP  \quad \bout \RR\msg\RS.B <: B' \quad
	\Xi=\Sigma \cup \funauth{\NA}{\Rnunauth\RR} \cup \funauth{\NA}{\Rnunauth\RS}
  \end{array}}
	\justifies
	\Delta\circ \NA:B'; \Gamma \vdash_{\Xi} \sauth{\NA}{\Rnunauth\RR}\msg{\Rnunauth\RS}.P
	 \endprooftree
  \\[8mm]
\mbox{[T-in]}  && 
  \prooftree
	 \Delta\circ \NA:B, \NB:B'; \Gamma \vdash_{\Sigma} P \quad \bin \RR\msg{B'}.B <: B'' \quad  \Xi=\Sigma \cup \funauth{\NA}{\Rnunauth\RR}
	 \quad \NB\not\in \channels{\Sigma}
	\justifies
	\Delta \circ \NA:B'' ; \Gamma \vdash_{\Xi} \receive\NA{\Rnunauth\RR} l\NB.P
	 \endprooftree
  \\[8mm]
  \mbox{[T-out]}  && 
  \prooftree
	 \Delta\circ \NA:B; \Gamma \vdash_{\Sigma} P \qquad \bout\RR \msg {B'}.B <: B'' \qquad \Xi=\Sigma \cup \funauth{\NA}{\Rnunauth\RR}
	\justifies
	\Delta \circ \NA:B'' \circ \NB:B'; \Gamma \vdash_{\Xi} \send\NA{\Rnunauth\RR} \msg\NB.P
\endprooftree
  \\[8mm]
 \mbox{[T-lsin]}  && 
  \prooftree
	\Delta\circ \NA:B'; \Gamma, \NB:T \vdash_{\Sigma} P \quad \bin\RR \msg T.B' <: B \quad \Xi=\Sigma \cup \funauth{\NA}{\Rnunauth\RR}
	\quad \NB\not\in \channels{\Sigma}
	\justifies
	\Delta \circ \NA:B ; \Gamma \vdash_{\Xi} \receive\NA{\Rnunauth\RR} \msg \NB.P
	 \endprooftree
  \\[8mm]
  \mbox{[T-lsout]}  && 
  \prooftree
	 \Delta\circ \NA:B'; \Gamma, \NB:T \vdash_{\Sigma} P \qquad \bout\RR\msg T.B' <: B \qquad \Xi=\Sigma \cup \funauth{\NA}{\Rnunauth\RR}
	\justifies
	\Delta \circ \NA:B; \Gamma, \NB:T \vdash_{\Xi} \send\NA{\Rnunauth\RR}\msg\NB.P
	 \endprooftree
  \\[8mm]
   \mbox{[T-sin]}  && 
  \prooftree
	\Delta, \NB:B; \Gamma, \NA:l(B) \vdash_{\Sigma} \PP \quad \NB\not\in \channels{\Sigma} 
	\justifies
	\Delta; \Gamma, \NA:\msg(B) \vdash_{\Sigma} \receive\NA{\Rauth{\RR}}\msg\NB.P
	 \endprooftree
% \\[8mm]
\quad \quad
  \mbox{[T-sout]}  \quad %&& 
  \prooftree
	 \Delta; \Gamma, \NA:\msg(B)\vdash_{\Sigma} P
	\justifies
	\Delta \circ \NB:B; \Gamma, \NA:\msg(B) \vdash_{\Sigma} \send\NA{\Rauth{\RR}}\msg\NB.P
	 \endprooftree
%  \\[8mm]
%  \mbox{[TProc-par]} && \prooftree
%	 \Delta_1; \Gamma \vdash_{\Sigma} \PP \qquad \Delta_2; \Gamma \vdash_{\Xi} \PQ
%	\justifies
%	\Delta_1 \circ \Delta_2; \Gamma \vdash_{\Sigma\cup\Xi} \PP \parop \PQ
%		 \endprooftree \\
%		 %\\\hline
\end{array}\]
%\hline\hline
\caption{Typing rules. We define: $
\funauth\NA{\Runauth\RQ}=\{(\NA,\RQ)\},\funauth\NA{\Rauth\RQ}=\emptyset$}
\label{tab:typing_rules}
}
\end{figure}

Typing rules are presented in Figure~\ref{tab:typing_rules}. There are two rules that are specific to our model:
\begin{enumerate}[-]
\item \rname{Trole-in} types authorization reception of the role $\RS$ on the linear channel $\NA,$ under the role $\Rnunauth\RR,$ with the authorization set diminished by $(\NA,\RS),$ and enlarged with $(\NA,\RR)$ in case $\Rnunauth\RR$ is $\Runauth{\RR}$.  
\item \rname{Trole-out}  types sending of the authorization $\Rnunauth\RS$ on linear channel $\NA,$ under the role $\Rnunauth\RR,$ with the authorization set enlarged with $(\NA,\RR)$ or $(\NA,\RS)$ in case $\Rnunauth\RR$ is $\Runauth{\RR}$ or $\Rnunauth\RS$ is $\Runauth{\RS}.$
\end{enumerate}
Notice that in both rules the typing environment in the conclusion is split in a typing of $\NA$ that specifies the reception of the authorization (up to subtyping).

All other rules are similar to the typing rules from~\cite{BaltazarCVV12}, with the derivation of the novel decoration $\Sigma$ as follows.
%\[1mm]
%\begin{itemize}
%\item[] 
Rule \rname{T-end} states that a well-typed inactive process has no unauthorized roles and only $\bend$ usages of linear channels (denoted by 
$\Delta_\bend$). %\\[1mm]
%\item[] 
Rule \rname{T-new} types a restricted linear name if it (a)~has no unauthorized roles and (b)~has no un{\textit{matched}} communications (no output or input communication prefixes). Rule \rname{T-snew} types a restricted shared name, without any additional restriction on unauthorized roles.
%\\[1mm]
%\item[] 
Rules \rname{T-in}, \rname{T-out}, \rname{T-lsin} and \rname{T-lsout} type input/output actions under the role $\Rnunauth\RR,$ with the authorization set enlarged  with 
$(\NA,\RR)$ in case $\Rnunauth\RR$ is $\Runauth{\RR}.$ Notice the typing environment in the conclusion of rules \rname{T-out} and \rname{T-sout}�
mentions the usage delegated in the communication (via splitting). %\\[1mm]
%\item[] 
Rules \rname{T-sin} and \rname{T-sout} state that input and output actions on shared channels are well typed only under authorized roles. The authorization is not performed on shared channels, implying that $\channels\Sigma$ are linear and not changed under actions on shared channels.
 %\\[1mm]
%\item[] 
Rule \rname{TProc-par} states that the unauthorized pairs in a parallel composition of two processes is the union of unauthorized pairs of the two composed processes.
%\end{itemize}
%
  We say that a process $\PP$ is \emph{well typed} if there are $\Delta$ and $\Gamma$ such that $\Delta; \Gamma \vdash_{\emptyset} \PP.$
\begin{proposition}[Error free]
 If $\PP$ is a well-typed process, then $\PP$ is not an authorization error.
\end{proposition}
We define the reduction relation $\red$ between behavioral types $B$ and corresponding environments $\Delta$ %(the rules are given in Figure~\ref{fig:EnvRed}) 
by allowing a synchronized communication prefix to reduce to its continuation $\btau{\RS}{\RR}\msg(M).B  \red B$, so as to mimic the respective 
process behavior, by allowing reduction to occur in a branch of a parallel composition (e.g., $B_1\red B_2 \implies B \;|\; B_1 \red B \;|\; B_2$),  
and by lifting the relation point-wise to environments, embedding reflexivity so as to encompass process reductions involving shared or bound channels 
(where no reduction in the linear usages of free names is required).
\begin{theorem}[Type Preservation]
  Let $\Delta;\Gamma\vdash_{\Sigma} \PP$ for some $\Delta, \Gamma, \Sigma$ and $\PP.$ 
  If $\PP \red \PQ$ then there is $\Delta'$ such that $\Delta \red \Delta'$ and $\Delta';\Gamma\vdash_{\Sigma} Q.$
\end{theorem}
A direct consequence of type preservation is  \emph{protocol fidelity}: every reduction of the process corresponds to a reduction of 
the types, thus ensuring that the process follows the protocols prescribed by the types. Notice that communication safety (no type
mismatches in communications) is entailed by protocol fidelity, which in our case also attests that processes agree in the role when
sending and receiving authorizations.
Combining freedom from errors and type preservation results we immediately obtain our notion of type safety, which ensures that well-typed processes never
reach an error configuration. 
\begin{corollary}[Type safety]\label{c:safety}
  If $\PP$ is a well-typed process and  $\PP\red^*\PQ,$ then $\PQ$ is not an authorization error. 
\end{corollary}
%
%\begin{corollary}[Conversation Fidelity]
%  Let $\Delta;\Gamma\vdash_{\emptyset} \PP$ for some $\Delta$ and $\Gamma.$ 
%  Then $\PP$ follows the protocols given by $\Delta$ and $\Gamma.$
%\end{corollary}
% !TEX root = main.tex
%========================
\section{Related Work and Concluding Remarks} 
\label{sec:related}
%=========================

%Security maintenance and privacy protection during transfer and management of data are important features in the design of distributed networks. These issues are reviewed for dynamic web documents handled by XML in~\cite{dezaghilpant06,dezaghilpantvara08,ghiljakspantdeza12}. 

Role-based access control in distributed systems with dynamic access rights was handled in~\cite{dezaghiljakspant11} by means of a type system, which ensures security properties. In this calculus, roles assigned to data can be dynamically administered, while role communication between processes was not treated.

Previous works consider security properties, like confidentiality and integrity, in the setting of session calculi.  For instance, in~\cite{BonelliCG05} session types are extended with correspondence assertions, a form of dependent types which ensures consistency of data during computation. More recently, aspects of secure information flow and access control %are controlled by security levels 
have been addressed for sessions in \cite{BonoCCD11,abs-1108-4465,CapecchiCDR10}. A kind of role-based approach is used in~\cite{DenielouY11}, where communication is controlled by a previously acquired reputation.

Similarly to our work, the work~\cite{Lapadula07} consider a typed approach to role-based authorizations, in the setting of service-oriented applications. Differently from our model, assigned roles are initially authorized and communicated data carries information on roles that will use it. 

Our contribution is based on the previous work on conversation types \cite{cairvier10} and their extension with dynamic assignment of roles to several parties in a concurrent system \cite{BaltazarCVV12}. We focused on a modular extension of the existing framework, so as to leverage on the previous results, adding the minimal elements so as to identify the specific issues at hand and set the basis for further exploration.
We consider the problem of role authorization and authorization passing in an extension of the $\pi$-calculus,  where communication prefixes are annotated with role authorizations. The underlying calculus allows for the dynamic communication of authorizations. We then extend the conversation type system in which a well-typed process can never incur in an authorization error. In this way we can statically distinguish processes that are always authorized to communicate on behalf of a role including when authorizations are dynamically passed in messages.
As a natural continuation of this study, we aim to extend the present calculus with tools that will enable role de-authorization. For this purpose, we aim to equip the type system with qualified (authorized or unauthorized) roles, instead of unqualified ones. In such a calculus, we could model authorization removal and authorization lending. Moreover, by introducing a partial order into the set of roles, we could control communicated roles with the aim to provide absence of authorization leaks.

%\vspace{0.2cm}
%\noindent
\paragraph{\textbf{Acknowledgments.}} 
We are grateful to the anonymous reviewers for their useful remarks.
This work was supported by COST Action IC1201: Behavioural Types for Reliable Large-Scale Software Systems (BETTY) 
%-- \url{http://www.behavioural-types.eu} -- 
via Short-Term Scientific Mission grants (to Pantovi\'c and Vieira), and by
FCT through project Liveness, PTDC/EIAÐCCO/117513/2010, and LaSIGE Strategic Project, PEstÐOE/EEI/UI0408/2014 and by grants ON174026 and III44006 of the Ministry of Education and Science, Serbia.

\bibliographystyle{eptcs}
\bibliography{main}

%\end{document}
%\newpage
%\appendix
%\input{appendix}
\end{document}